\documentclass[9pt,a4paper]{article}

\usepackage[a4paper,top=18mm,bottom=19mm,left=18mm,right=18mm,headsep=6mm,footskip=9mm]{geometry}
\usepackage{fontspec}
\defaultfontfeatures{Ligatures=TeX,Scale=MatchLowercase}
\usepackage{microtype}
\usepackage{graphicx}
\usepackage{xcolor}
\usepackage{multicol}
\usepackage{longtable,booktabs,array,calc,multirow,tabularx}
\usepackage{etoolbox}
\usepackage{ragged2e}
\usepackage{fancyhdr}
\usepackage{hyperref}
\usepackage{xurl}
\usepackage{bookmark}
\usepackage{footnotehyper}

\definecolor{JournalBlue}{HTML}{173B57}
\definecolor{JournalCyan}{HTML}{2C7DA0}
\definecolor{JournalPale}{HTML}{EEF4F7}
\definecolor{JournalText}{HTML}{22272B}
\definecolor{JournalGray}{HTML}{5D6870}
\color{JournalText}

\hypersetup{
  colorlinks=true,
  linkcolor=JournalBlue,
  urlcolor=JournalCyan,
  citecolor=JournalBlue,
  pdftitle={De mora luminis: Roemer's discovery 350 years later},
  pdfauthor={Fabio Falchi, Riccardo Furgoni, Paolo Gattillo, Maurizio Francesio},
  pdfcreator={Professional journal layout}
}
\raggedcolumns
\makeatletter
\patchcmd\longtable{\par}{\if@noskipsec\mbox{}\fi\par}{}{}
\makeatother
\makesavenoteenv{longtable}

\makeatletter
\def\maxwidth{\ifdim\Gin@nat@width>\linewidth\linewidth\else\Gin@nat@width\fi}
\def\maxheight{\ifdim\Gin@nat@height>.80\textheight .80\textheight\else\Gin@nat@height\fi}
\makeatother
\setkeys{Gin}{width=\maxwidth,height=\maxheight,keepaspectratio}

\renewcommand{\headrulewidth}{0.35pt}
\renewcommand{\headrule}{\hbox to\headwidth{\color{JournalCyan}\leaders\hrule height \headrulewidth\hfill}}

\renewcommand{\footnoterule}{\kern-2pt{\color{JournalCyan}\hrule width 18mm height .45pt}\kern4pt}

\makeatletter
\renewcommand\section{\@startsection{section}{1}{0pt}%
  {1.15\baselineskip plus .2\baselineskip minus .1\baselineskip}%
  {.52\baselineskip}%
  {\sffamily\bfseries\fontsize{10.4}{12}\selectfont\color{JournalBlue}\MakeUppercase}}
\renewcommand\subsection{\@startsection{subsection}{2}{0pt}%
  {.85\baselineskip plus .15\baselineskip minus .1\baselineskip}%
  {.35\baselineskip}%
  {\sffamily\bfseries\fontsize{9.3}{10.7}\selectfont\color{JournalBlue}}}
\makeatother
\newenvironment{titlezone}
  {\thispagestyle{fancy}\noindent{\color{JournalCyan}\rule{\textwidth}{1.4pt}}\par\vspace{1.5mm}%
   {\sffamily\fontsize{6.9}{8.2}\selectfont\color{JournalGray}\RaggedLeft
   This is an author-formatted text of the paper published on the European Journal of Physics, 47 (2026) 025802
   \href{https://doi.org/10.1088/1361-6404/ae3f66}{https://doi.org/10.1088/1361-6404/ae3f66}\par}\vspace{5.5mm}}
  {\vspace{2.5mm}\noindent{\color{JournalCyan}\rule{\textwidth}{.55pt}}\par\vspace{4mm}}
\newenvironment{journaltitle}
  {\sffamily\bfseries\fontsize{21}{24}\selectfont\color{JournalBlue}\RaggedRight}
  {\par\vspace{5mm}}
\newenvironment{journalauthors}
  {\sffamily\bfseries\fontsize{10}{12}\selectfont\color{JournalText}\RaggedRight}
  {\par\vspace{2.5mm}}
\newenvironment{affiliations}
  {\sffamily\fontsize{7.9}{9.4}\selectfont\color{JournalGray}\RaggedRight\setlength{\parskip}{0pt}}
  {\par\vspace{1.5mm}}
\newenvironment{deceasednote}
  {\sffamily\fontsize{7.6}{9}\selectfont\color{JournalGray}\RaggedRight}
  {\par\vspace{1.2mm}}
\newenvironment{correspondence}
  {\sffamily\fontsize{8}{9.5}\selectfont\color{JournalGray}\RaggedRight}
  {\par\vspace{3mm}}
\newcommand{\journalfrontlabel}[1]{%
  {\sffamily\bfseries\fontsize{8.5}{10}\selectfont\color{JournalBlue}\MakeUppercase{#1}}\par\vspace{1mm}}
\newenvironment{contributions}
  {\noindent\begin{minipage}{\textwidth}\sffamily\fontsize{7.8}{9.4}\selectfont\color{JournalGray}}
  {\end{minipage}\par\vspace{3mm}}
\newenvironment{abstractblock}
  {\noindent{\color{JournalCyan}\rule{\textwidth}{.55pt}}\par\vspace{2mm}\noindent\begin{minipage}{\textwidth}\fontsize{9.1}{11.2}\selectfont\setlength{\parindent}{0pt}\setlength{\parskip}{3pt}}
  {\end{minipage}\par\vspace{2mm}\noindent{\color{JournalCyan}\rule{\textwidth}{.55pt}}\par}

\newenvironment{widetable}
  {\par\medskip\begingroup\fontsize{7.5}{9}\selectfont\setlength{\tabcolsep}{3.2pt}}
  {\par\endgroup\medskip}
\newenvironment{widefigure}
  {\par\medskip\begingroup\centering}
  {\par\endgroup\medskip}
\newenvironment{journalcaption}
  {\par\smallskip\begin{minipage}{\textwidth}\sffamily\fontsize{7.6}{9.3}\selectfont\color{JournalGray}\RaggedRight}
  {\end{minipage}\par}

\ExplSyntaxOn
\cs_new_protected:Npn \journal_cite_number:n #1
  {
    \hypertarget{cite-#1}{}
    \hyperlink{ref-#1}{\textcolor{JournalBlue}{\textbf{#1}}}
  }
\NewDocumentCommand{\journalcite}{m}
  {
    \textcolor{JournalBlue}{\textbf{[}}
    \group_begin:
    \bool_set_true:N \l_tmpa_bool
    \clist_map_inline:nn {#1}
      {
        \bool_if:NF \l_tmpa_bool
          {\textcolor{JournalBlue}{\textbf{,\c_space_tl}}}
        \journal_cite_number:n {##1}
        \bool_set_false:N \l_tmpa_bool
      }
    \group_end:
    \textcolor{JournalBlue}{\textbf{]}}
  }
\ExplSyntaxOff
\newenvironment{journalreferences}
  {\begingroup\fontsize{7.45}{9.1}\selectfont
   \interlinepenalty=10000
   \begin{list}{}{\setlength{\leftmargin}{5.2mm}%
     \setlength{\labelwidth}{4.6mm}\setlength{\labelsep}{.6mm}%
     \setlength{\itemsep}{2.2pt}\setlength{\parsep}{0pt}%
     \setlength{\topsep}{2pt}}}
  {\end{list}\endgroup}

\begin{document}
\fontsize{8.65}{10.45}\selectfont
\begin{titlezone}

\begin{journaltitle}

De mora luminis: Roemer's discovery 350 years later

\end{journaltitle}

\begin{journalauthors}

\textbf{Fabio Falchi}\textsuperscript{1,2,3*}, \textbf{Riccardo
Furgoni}\textsuperscript{4}, \textbf{Paolo Gattillo}\textsuperscript{5}
and \textbf{Maurizio Francesio}\textsuperscript{6,\textsuperscript{†}}

\end{journalauthors}

\begin{affiliations}

\textsuperscript{1} ICA, Slovak Academy of Sciences, Bratislava,
Slovakia.

\textsuperscript{2} Istituto Superiore `E. Fermi', Mantova, Italy

\textsuperscript{3} ISTIL, Light Pollution Science and Technology
Institute, Thiene, Italy

\textsuperscript{4} Fondazione GAL Hassin - Centro Internazionale per le
Scienze Astronomiche, Isnello, Italy

\textsuperscript{5} CieloBuio, Associazione per la protezione del cielo
notturno, Varese, Italy

\textsuperscript{6} Retired professor of Physics, Mantova, Italy

\end{affiliations}

\begin{deceasednote}

\textsuperscript{†} Deceased.

\end{deceasednote}

\begin{correspondence}

*Corresponding author:
\href{mailto:falchi@istil.it}{\emph{falchi@istil.it}}

\end{correspondence}

\begin{contributions}\journalfrontlabel{Author Contributions}

Conceptualization: FF, MF; Methodology: FF, MF; Investigation: FF, RF,
PG; Formal Analysis: FF; Visualization: FF, RF; Supervision: FF,
Writing---original draft: FF; Writing---review \& editing: FF, RF, PG.

\end{contributions}

\begin{abstractblock}\bfseries\journalfrontlabel{Abstract}

\textbf{350 years from the 1676 announcement of the Roemer's discovery
that light propagates with finite speed, we present our observations
using eyes with telescopes having similar resolution compared to those
in the late 17th century. We confirmed that Roemer's method is valid and
gives reasonable values for the speed of light
}\emph{\textbf{c}}\textbf{, within about 10\% of the modern value for
our measurements, even with the simplest modelling technique, using
uniform circular motions. We found that increasing the complexity of the
model, e.g., by taking into account the elliptical orbit of Jupiter,
does not necessarily bring the results closer to the value of
}\emph{\textbf{c}}\textbf{ due to the influence of other perturbations.
Using modern ephemerides yields a noticeably accurate result of
c=(298200±1900) km s}\textsuperscript{\textbf{-1}}\textbf{.}

This experience can have great didactic value by showing the
interconnections between formulation of hypotheses and the consequent
predictions, making observations, reducing data, and searching for
alternative explanations for the same phenomenon.

Lastly, we also found, in the correspondence between Roemer and Huygens,
that Roemer in 1677 searched for an independent confirmation of what he
found during previous years observing Io's eclipses by making
observations and reducing the data of the meridian transits of the Great
Red Spot on Jupiter.

\end{abstractblock}\end{titlezone}

\setcounter{footnote}{0}\begin{multicols}{2}

\section{INTRODUCTION}

Ole Roemer's method to calculate the speed of light is a formidable way
to understand how science works, and thus has a great didactic value,
ideally aimed at undergraduate level.

In the 1670's several factors converged toward the possibility of
measuring the speed of light: the first precise measurement of Solar
System dimensions (1673), the availability of precise pendulum clocks,
sufficiently good telescopes, and the forecast of future Jupiter
satellites events. Roemer himself seems to not have given any figure for
the speed of light, but the fundamental thing was that he demonstrated
that light travels at a finite speed, now denoted\emph{ c},\emph{ }and
consequently its transit is not instantaneous, as most of the natural
philosophers, following Descartes, thought at that time.

Now 350 years have passed since those days and more than 40 since one of
us, F.F., was introduced to this fascinating theme by reading an article
published by his high school physics teacher, Maurizio Francesio, in the
journal of AIF, the Italian Association for Physics Teaching~\journalcite{1}. Guided by the article and directly by the teacher, the
student, then at the second year of Liceo Scientifico using his 114 mm
reflector telescope, timed a few eclipse ends and reduced the data,
obtaining two different values, 200000 and 400000 km
s\textsuperscript{-1}. These original observations are apparently lost.

It remained a curiosity to directly verify if the method is viable to
obtain a reasonable figure for the speed of light. We decided to repeat
the observations using the eye and small telescopes, comparable in
diameter to those used by Roemer, and using also a digital camera
applied to another relatively small telescope similar to those often
available in colleges.

In this paper, Section I gives a brief historic context of the Roemer
method; Section II describes the observations; and Section III describe
the models and data reduction to obtain the measure of the speed of
light using different assumptions, discussing the discrepancies with the
accepted value of \emph{c} and their possible causes. Section IV
outlines the didactic value and draws our conclusions.

\section{I HISTORICAL CONTEXT}

\subsection{\texorpdfstring{ Galileo \emph{Gedankenexperiment} to
measure the speed of
light}{ Galileo Gedankenexperiment to measure the speed of light}}

Galileo Galilei was the first to try to measure the speed of
light~\journalcite{2}, in case its propagation wasn't instantaneous as most
then thought. He proposed an experiment where each of two observers hide
and show a light with their hand. When the first shows his light, the
second immediately shows his, too, and the first judges if he sees the
other's light immediately after he has shown his. After testing at close
distance what we call now reaction times~\journalcite{3}, the
observers would be relocated at a certain distance, three miles (or even
10 by using telescopes to observe each other), and repeat the
experiment. Galileo made the experiment, but at a distance of slightly
less a mile, and found no detectable delay compared to the experiment
made at very close distance. This can give an estimation of a lower
limit for the speed of light. Supposing a distance of 1 km and a
reaction time of 0.2 s, the speed of light is greater than 10 km
s\textsuperscript{-1}, i.e., at least 30 times faster than the speed of
sound.

\subsection{\texorpdfstring{ Descartes' arguments for the instantaneous
propagation of
light}{ Descartes' arguments for the instantaneous propagation of light}}

Paradoxically, it was easier to accept an instantaneous propagation of
light, i.e., to assume an infinite speed, instead of accepting a merely
very high speed. Descartes, for example, used the argument that the
speed of light cannot be finite because otherwise during a lunar eclipse
the Sun, Earth and the Moon would not be in a straight line. In fact, he
assumed an absurdly high velocity, for his times, of one hour for light
to travel the Earth-Moon distance and deduced that by the time the light
travelled this distance during an eclipse the Moon would no longer be in
perfect alignment with the Sun. Christiaan Huygens counter-argued, in
his letter to Roemer of September 16, 1677~\journalcite{4}, that if the transit time
needed by the light would have been only 2 seconds, consistent with
Roemer's data he read in Philosophical Transactions,~\journalcite{5} and further if the Earth-Sun
distance equaled 12000 Earth diameters, no misalignment would be
perceivable.

\subsection{\texorpdfstring{ Telescopes and clocks in second half of the
17th
century}{ Telescopes and clocks in second half of the 17th century}}

Roemer and his colleagues had at their disposal telescopes that, having
only one objective lens and therefore suffering from chromatic
aberration, were of great quality, vastly outperforming those available
in the first half of the century. The best opticians were able to
produce high-performing lenses due to the innovations in lens making and
improvement in glass
homogeneity~\journalcite{6, 7}. The best telescopes allowed observers to
see very fine planetary details, as testified by the discovery of the
major division in Saturn's rings by Gian Domenico Cassini~\journalcite{8} using a 108-mm-diameter refractor made by the
optician Giuseppe Campani. The possibility to see what is now called
Cassini's Division in the rings is validated by interferometric tests of
historically made lenses that survived to our time~\journalcite{9}.

With the idea of using pendula to regulate mechanical clocks by Galileo
and the development by Huygens of real pendulum clocks in 1657 and
subsequent years, sufficiently precise timekeeping became possible,
reducing the daily timing error below 15 seconds~\journalcite{10}.

\subsection{\texorpdfstring{ Dimensions of Earth and the Solar
System}{ Dimensions of Earth and the Solar System}}

The Earth's dimensions were still not very well known in 17th century
such that the French Académie Royale des Sciences asked for a new
measurement, carried out by Jean Picard. He found a value of 57060
toises (of the Châtelet de Paris) for the length of one degree of
meridian arc~\journalcite{11}, corresponding to 40036 km for the meridional
circumference of Earth, improving the precision of the value found by
Eratosthenes almost 2000 years earlier (252000 stadia, corresponding to
39690 km using a stadium length of 157.5
m)~\journalcite{12, 13, 14}. At the time of
Picard, the measure of Eratosthenes was not well known because of the
uncertainty on which of the different stadia lengths in use during his
time that he adopted for his 252000 figure\footnote{It has also been plausibly suggested \hyperlink{ref-14}{\textcolor{JournalBlue}{\textbf{[14]}}} that the `Eratosthenic
  stadium' was specifically defined as 1/252,000 of the
  Earth\textquotesingle s circumference, in a manner analogous to the
  later definition of the meter.} and, moreover, modern measurements made in the early part
of 17\textsuperscript{th} century gave different circumference values,
from 38660 km by Snell and 40200 km by Norwood.

At the time, the distances between planets of our solar system involved
much higher uncertainties. The relative distances were well known, but
their absolute values weren't. To determine the distances between the
planets, Cassini planned simultaneous observations of the position of
Mars observed from Paris and from Cayenne, French Guiana, where Jean
Richer was sent in the years 1672-1673. The measured parallax gave the
first reliable values of distances inside the Solar System,
underestimating the modern values by about 8\%, giving a value for the
Sun-Earth distance, the Astronomical Unit (A.U.), of 21600 times the
Earth radius, corresponding to about 137.6 million km using the meridian
value found by Picard.

\subsection{\texorpdfstring{ The longitude
problem}{ The longitude problem}}

In 17th century determining one's longitude on Earth was an unsolved
problem, both on land but especially also at sea. Latitude was well
determined simply by measuring the altitude above the horizon of the
star Polaris or of the Sun. Longitude is much harder to determine, as
the appearance of the sky for sites at a given latitude, is identical,
once the Earth rotates by an angle equal to the longitude difference of
two sites. For example, consider the night sky seen from both Naples and
New York, two cities approximately on the same parallel of latitude. We
now know that New York is 88° to the west of Naples, so we expect to see
any given celestial object transit the local meridian in Napes 5 hours
and 52 minutes before it transits in New York, accounting for the
rotation of the Earth by 88° during that time interval.

A way to find the longitude would be to determine the instant when a
star culminates, i.e., passing over the exact south while crossing the
celestial meridian. At those times this could be done, but only with the
local time, that depends on longitude. We therefore need a way to
synchronize clocks. This can be done by observing the same lunar eclipse
in two different places. For example, the start of an eclipse
simultaneously visible in Naples and New York would be seen at two
different local times, say at 4:00 a.m. in Naples and at 10:08 p.m. (of
the previous day's date) in New York. However, there are some problems
with this approach: lunar eclipses are rare (and so are useful only for
terrestrial locations, and not to find the position of a ship when
needed), and the exact timing of the start or end of an eclipse is
uncertain due to the fact that Earth's shadow seen falling on the Moon
is insufficiently sharp. Better astronomical events can be used for this
purpose, and Galileo found them in the eclipses, occultations and
transits of the four main satellites of Jupiter. In his words
(translation by us):

``\emph{\emph{Then, again by means of ephemerides calculated by me
hourly, in which the moments of the said conjunctions, separations, and
eclipses are contained for long periods to come, one will come, during
the same navigation, at any hour of the night, to certainty of the true
longitude, and consequently of the true location where the ship is
located; and this for ten months of each year, it being true that for
two months at most these new stars}}\footnote{Galileo is referring to Jupiter satellites.}\emph{\emph{ remain invisible, which is when the Sun is
close to them.}}''

This is an excerpt from the General Report on Galileo Galilei's new
invention for determining longitude at any time or place (in Italian:
\emph{Relazione Generale del nuovo trovato di Galileo Galilei in
proposito del prendere in ogni tempo e luogo la longitudine)
}~\journalcite{15}. Galileo refers to this as the "general explanation of my
invention" in a letter to Orso D'Elci---the Grand Duke of Tuscany's
ambassador to Spain---dated November 13, 1616~\journalcite{16}. Galileo had already proposed the method in 1612, albeit
with fewer details so as to safeguard his claim to its
invention~\journalcite{17}.

The problem to find longitude at sea and on land was one of the main
drivers in contemporaneous observations of the Galilean satellites. In
fact, Picard went to Uraniborg on Hven island, then part of Denmark, to
find the longitude difference with Paris. To determine it, he used
timings of Galilean satellites events observed at the same time in the
two locations. The difference in local times at which events were
observed indicated the difference in longitude between Paris and
Uraniborg. Picard was helped by an extraordinary observer, Ole
Christensen Roemer. Picard asked Cassini to hire this young man for work
at the Paris Observatory. In fact, Roemer accompanied Picard on his
return to Paris and worked in France for 10 years, where he continued
observing the Jupiter satellites events that resulted in the discovery
that light does not propagate instantaneously.

\subsection{Roemer's method to determine the speed of light}

Roemer's method to prove the finite speed of light was very well
described by himself in the note on Journal des Sçavans~\journalcite{18} published on December 7, 1676. Taking into account the
observations made at the Académie Royale des Sciences and at the
Observatory of Paris in the previous eight years he found that when
Earth was approaching Jupiter (i.e., between Jupiter's conjunction with
the Sun and opposition), the periods between successive eclipses of the
satellite Io were shorter than the average and the contrary happened
when Earth was receding from Jupiter (i.e., after opposition until
conjunction). This shortening or delay in the periods could not be
measured during a single revolution of Io around Jupiter due to
uncertainty in time measurements, but it became evident over a large
number of revolutions: in Roemer's example, 40 orbits which take about
71 days to complete. In this longer period, the anticipation or delay
was clearly measurable.

Roemer explained this phenomenon with the hypothesis that the speed of
light is finite. Indeed, when the Earth is receding, the actual time
interval between the end of two eclipses---as might be measured on Io
itself---appears delayed to an observer on Earth by the time required
for light to travel the accumulated additional distance between Jupiter
and the receding Earth. Roemer gave a total time of 22 minutes for the
light to cross the entire diameter of Earth's orbit around the Sun.

Figure 1, panel a, explains the geometry of the situation, with the
dates of our first and last observations, taken while Io orbited 70
times around Jupiter. The space travelled by light in the last
observation is much greater than that in the first, near opposition when
the two planets were closer, as is apparent by comparing the length of
the two solid black lines. Panels b and c show the satellite Io while
exiting the planet's shadow.

\subsection{Cassini's critics}

The Director of Paris Observatory, Gian Domenico Cassini, at first
favoured this explanation that perhaps was made by him for the first
time, as described by Bobis and Lequeux~\journalcite{19}. Later
Cassini refuted it, mainly because the anticipations and delays of the
Io eclipse times were not so clear for the other three satellites, and
Cassini himself gave full credit to Roemer for the idea that these
irregularities were caused by the finite speed of light~\journalcite{20}.

In a long letter to Christiaan Huygens~\journalcite{21} dated September 30, 1677,
Romer explains that the three outer satellites were not well suited to
obtain data on the speed of light because for them (translation by us):

I. Immersions and emersions are rarer.

II. The moments of contact of the shadow are less precise, both because
of their slower movement and because they mostly fall obliquely on the
periphery of the shadow\footnote{Also because the penumbral region within Jupiter's shadow is
  necessarily wider, given the greater distance of these three outermost
  satellites from the planet.}.

III. Their inclinations and nodes are not well known so that there is a
discrepancy of many minutes in the oblique incidences of the shadow.

IV. It is admitted that they have irregularities that have not yet been
determined, whether it is eccentricity or some other cause that makes
observations deviate from the theories of D. Cassini, over a period of
time twice or three times greater than that which we are investigating
here and determining from the first Satellite.

In the same letter, Roemer envisions the possibility of using the
rotation of Jupiter, by observing the then-newly discovered Great Red
Spot, to verify its apparent acceleration or slowing down due to the
delay of light in reaching the observer. Of course, this is an
ante-litteram Doppler effect, as already noted
elsewhere~\journalcite{22, 23}. We found that, in a memoire
presented to the Académie Royale des Sciences on December 18,
1677~\journalcite{24}, Roemer
reported his and Cassini's observations and timing of the Great Red Spot
transits on Jupiter's central meridian taken on September 7 and 12 and
December 8, 1677. From these observations, using the calculation of the
number of rotations of Jupiter made by Cassini, Roemer found that the
time needed by light to cross the additional space due to the increase
of the Jupiter-Earth distance, corresponding to 1¼ Earth-orbital
radii\footnote{We found a value 1.235 A.U. using modern ephemerides.
  Roemer probably used, in his written memory, the nearest fraction, 1¼,
  that approximated well the value he found.}, was 14 minutes. This time
corresponds to 22.4 minutes for the diameter of Earth's orbit, nearly
confirming what he already published the previous year. In the same
memoire, Roemer reported also new observations using Io's eclipses,
giving a retarded time of 12 minutes for light travelling the same 1.25
radii of Earth's orbit, i.e., slightly more than 19 minutes for the
orbit diameter. As a point of comparison, modern measurements show that
to cross Earth's orbit, light needs 16.6 minutes.

It is to be noted that, while Roemer gives times for travelling
distances expressed in units of Earth's orbit radii\footnote{That is, what we call the Astronomical Unit (A.U).}, as far as we know, he never gave a figure for the speed
of light, while yet proving its finite nature. With the Sun-Earth
distance found in 1672 by Cassini, Picard and Richer, the speed of
light, obtained by the above-mentioned Roemer's time measurements,
varies from 242000 to 283000 km s\textsuperscript{-1}. The modern figure
is 299792 km s\textsuperscript{-1} in vacuum.

\end{multicols}

\begin{widefigure}

\includegraphics[width=17cm,height=12.478cm]{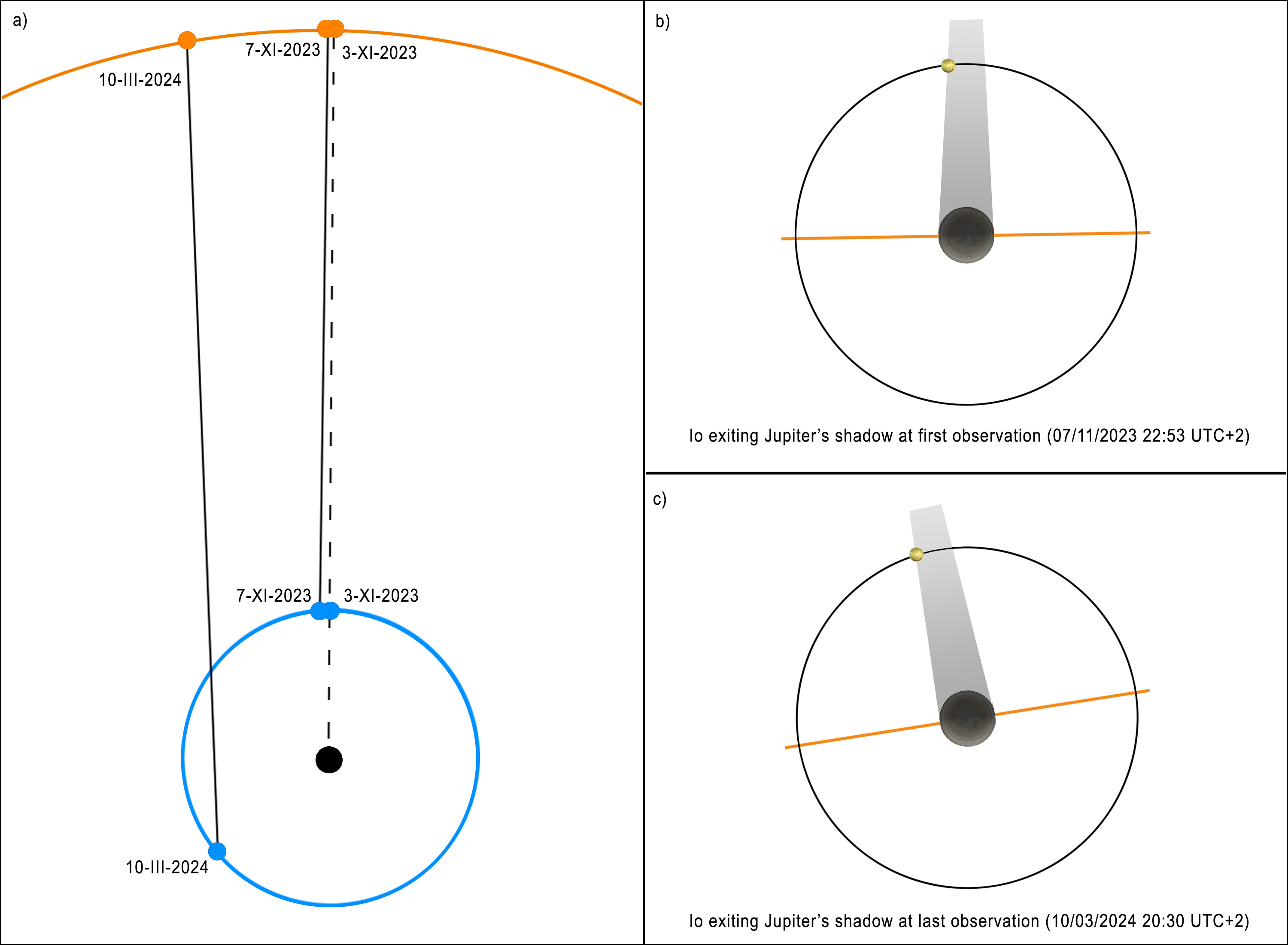}

\begin{journalcaption}

Figure 1. Panel a: Earth's orbit and positions of our planet (in blue)
on the given dates (opposition, first and last observation of Io's
eclipses) and Jupiter's positions on the same dates (in orange). The
dashed lines connect the Sun (black), Earth and Jupiter at opposition on
November 3, 2023. The solid lines illustrate the Earth-Jupiter distances
at the time of first and last observations. Orbits are shown to scale,
but celestial bodies are not shown to scale.

Panel b: emersion of Io from Jupiter's shadow on November 7, 2023. Panel
c: emersion of Io from Jupiter's shadow on March 10, 2024. Planet
Jupiter and Io's orbit are shown to scale, but the size of Io is not
shown to scale. In panels b and c, the orange line of Jupiter's orbit is
inclined in respect with inclination zero (horizontal in the panels) at
opposition.

\end{journalcaption}\end{widefigure}\begin{multicols}{2}

\section{II OUR OBSERVATIONS}

Two of us, FF and PG, observed with the eye using small refracting
telescopes and one, RF, used a 203 mm reflector coupled with a SLR
digital camera. Details are in Table 1.

The refracting telescopes we used had diameters comparable or slightly
smaller than those used by Cassini and his troupe at Paris Observatory.
The 80 mm was on an altazimuth mount that required users to manually
rotate two knobs to keep the object (Jupiter) inside the field of view,
simulating somewhat the observing conditions of 17th century. The 115 mm
telescope had a motorized equatorial mount. The Cassini Division is
barely visible with the 80 mm telescope and evident with the 115mm one,
confirming that---for our purposes---their optical performance is
comparable to that of the telescopes used in the late 17th century.

The reflector used to collect the digital images was similarly mounted
equatorially and could automatically follow the motion of the planet in
the sky. Being of a larger aperture than the two refractors, this
telescope has a better resolution and larger collecting area. This,
combined with the sensitivity of the digital camera, may have an
advantage in detecting earlier Jupiter's satellites emerging from the
shadow of the planet. This could translate into better precision of the
measurement, but not necessarily if the time offset with respect to eye
observations is relatively constant.

\end{multicols}

\vfill
\begin{widetable}

\begin{journalcaption}

Table 1. Observations' settings

\end{journalcaption}

\begin{longtable}[]{@{}lllllll@{}}
\toprule\noalign{}
Observer & Telescope type & Diameter (mm) & Focal length (mm) &
Magnification (times) & Sensor & Timing \\
\midrule\noalign{}
\endhead
\bottomrule\noalign{}
\endlastfoot
Falchi & Fluorite doublet Refractor & 80 & 555 & 150 & Eye &
stopwatch \\
Furgoni & Schmidt-Cassegrain & 203 & 2030 & - & CMOS & Camera time \\
Gattillo & Triplet APO Refractor & 115 & 800 & 114 & Eye & stopwatch \\
\end{longtable}

\end{widetable}
\newpage
\begin{multicols}{2}

The time used for determining the instant of emersion was the atomic
time given by Atomic Clock \& Watch Accuracy Tool and ClockSync Android
Apps, checked with a radio-controlled watch receiving the signal from
the Frankfurt Atomic Clock.

For the eye observations, the time was transferred to an Omega
Speedmaster chronograph used as a stopwatch with a difference of less
than half a second. This allowed observers to operate in the dark,
keeping the eye constantly at the eyepiece and stop the time at the
first hint of reappearance of the satellite. To avoid false stops that
may have compromised the timing, a mental count was started at the first
hint of sight and, reaching certainty, the clock was stopped. This
happened three to five seconds later. Subtracting these delays from the
timing of observations does not increase the precision of the dataset.

For the digital observation data, the camera was set manually using the
above-mentioned atomic time. A few minutes before the forecasted time of
the satellite's emersion from Jupiter's shadow the camera started taking
pictures at a rate of one every 3 seconds. To determine the instant of
reappearance after eclipse, the observer flagged the first image that
unambiguously showed the illuminated satellite and noted the time at
which the image was obtained. An alternative detection time was
determined using the light curve of emersion, choosing the time when the
brightness of the satellite reached half of its maximum value, i.e. when
it is fully outside the shadow of Jupiter. This second method, while
apparently more robust, was found to be less precise than the time of
the first frame when the satellite became visible. In fact, the
differences between the free open source planetarium software Stellarium
(version 0.19.1;
\href{https://stellarium.org/}{\emph{https://stellarium.org}}) times of
emersions and the observed ones have a standard deviation of 3 seconds
with the first, simpler method and 7 seconds with the light curve
method. Also, the R\textsuperscript{2} value of the linear regression
graphs of distance over delay in emersion times have a value of 0.99998
for the first method and 0.994 for the second. Probably using synthetic
light curves to fit the data, as made by Mallama et al.~\journalcite{25}, could have given a slight advantage over the simple
method, but was beyond the aims of our work.

In Table 2 we report our observations, along with the timings by modern
ephemerides. These last timings were determined with Stellarium by
observing the monitor at the maximum enlargement in a darkened room, and
taking the end of eclipse when the disc of the satellite just started to
be directly illuminated by sunlight, showing the very first `dent' of
light in the edge that was emerging from Jupiter\textquotesingle s
shadow. Timing flagged with asterisks in Table 2 were not used, as
compromised by variable cloud presence. We note that the timings taken
by observing with the eye, by FF and PG, have a slight delay compared to
the values given by Stellarium. The delay is lower for the bigger
telescope that collects about twice the light of the smaller instrument,
allowing an earlier detection. The timing taken with a CMOS-equipped
digital camera resulted in a half-minute earlier, compared to the values
from Stellarium. This is probably due to how the shadow of the planet is
calculated by the software. As the Sun's apparent diameter, as seen from
Jupiter, is about 0.1°, the satellite takes about 43 seconds to cross
the penumbral width. Moreover, the refraction of sunlight in the high
atmosphere of Jupiter anticipates the real emersion of Io (Mallama et
al. Cit.). The standard deviation of the eye's timings are about 12-13
seconds, while that of the CMOS timings is 3 seconds (see the next
section for a brief discussion).

Observer FF collected eight observations, with 70 Io orbits from the
first to the last, and PG and RF made five each, but one each was
discarded for the presence of clouds. In fact, the detected times of
emersions were more than half a minute late, compared to the average
difference of the forecast time from Stellarium. PG's observations
covered 53 orbits while RF's were of 61 orbits. The emersions were
usefully observed on 10 different dates, six of which were monitored
simultaneously by two of us. No single event was observed by all three
observers.

\end{multicols}\begin{widetable}

\begin{journalcaption}

Table 2. List of the observations. The ``Stellarium'' column lists the
timing of Io's emersion from Jupiter's shadow obtained using this
planetarium software. The following three columns list the timing
obtained by each observer. The observations with asterisks were excluded
from the computation due to variable presence of clouds. The last line
reports the average difference between the time detections by each
observer and those predicted by Stellarium, along with the standard
deviation of the difference.

\end{journalcaption}

\begin{center}
\begin{tabular}{@{}
  >{\raggedright\arraybackslash}p{26mm}
  >{\centering\arraybackslash}p{22mm}
  >{\centering\arraybackslash}p{20mm}
  >{\centering\arraybackslash}p{20mm}
  >{\centering\arraybackslash}p{22mm}
  >{\raggedright\arraybackslash}p{50mm}@{}}
\multirow{2}{=}{Date} &
\multicolumn{4}{c}{Emersion time (UTC+2; hh:mm:ss)} &
\multirow{2}{=}{Notes} \\
\cmidrule(lr){2-5}
& Stellarium & FF (eye) & PG (eye) & RF (CMOS) & \\
\midrule
07/11/2023 & 22:53:42 & 22:53:40 & & & \\
15/11/2023 & 0:48:23 & & 0:48:13 & & \\
16/11/2023 & 19:16:59 & & & 19:17:10* & RF: clouds disturbed timing \\
23/11/2023 & 21:11:48 & 21:11:35 & & 21:11:11 & \\
08/12/2023 & 1:01:40 & & 1:02:18* & & PG: clouds disturbed timing \\
16/12/2023 & 21:25:35 & 21:25:32 & & 21:24:58 & \\
23/12/2023 & 23:20:50 & 23:21:00 & 23:20:42 & & FF: slight haze/veils \\
25/12/2023 & 17:49:31 & 17:49:29 & & 17:48:59 & \\
15/01/2024 & 23:35:35 & 23:36:03 & 23:35:56 & & FF: slight haze/veils \\
24/01/2024 & 19:59:40 & 19:59:48 & & & FF: slight haze/veils \\
16/02/2024 & 20:14:52 & & 20:14:58 & & \\
10/03/2024 & 20:29:45 & 20:30:06 & & 20:29:13 & FF: bad seeing; pointing
issue \\
Difference with Stellarium (s) & 0 (reference) & 6 ± 14 & 2 ± 14 & -35 ±
3 & Negative value indicates a detection before the Stellarium timing \\
\bottomrule
\end{tabular}
\end{center}

\end{widetable}\begin{multicols}{2}

\section{III MODEL AND DATA REDUCTION}

\subsection{Uniform circular model}

The first aim of the project was to test if a simple model based on
uniform circular motions could give reasonable speed of light
determinations. For this purpose we used as radii of the circular orbits
the now-accepted values of the semimajor axis of the elliptical orbits
of Earth and Jupiter and the corresponding sidereal periods to calculate
the angular orbital velocities \emph{w} (see Table 3).

\end{multicols}\begin{widetable}

\begin{journalcaption}

Table 3. Earth and Jupiter orbital characteristics used in this work.

\end{journalcaption}

\begin{center}
\begin{tabular*}{\textwidth}{@{\extracolsep{\fill}}llllll@{}}
\toprule
\multicolumn{3}{c}{Earth} & \multicolumn{3}{c}{Jupiter} \\
\cmidrule(lr){1-3}\cmidrule(lr){4-6}
Sidereal year (days) & Orbital radius (km) & Angular velocity w (rad/s)
& Sidereal period (days) & Orbital radius (km) & Angular velocity w
(rad/s) \\
\midrule
365.2564 & 149597870 & 1.9909864x10\textsuperscript{-7} & 4332.589 &
778479000 & 1.6784895x10\textsuperscript{-8} \\
\bottomrule
\end{tabular*}
\end{center}

\end{widetable}\begin{multicols}{2}

We computed, with a uniform circular model, the Earth-Jupiter distances
over the period of observations, from November 7, 2023 (slightly less
than 5 days after opposition) to March 10, 2024. We assumed as `real'
the distances given by Stellarium using the VSOP87 planetary solution
ephemerides~\journalcite{26}. There was an offset
between our simple model and these ephemerides of about 30 million
kilometres due to the fact that Jupiter was relatively near perihelion,
having passed it on January 20, 2023. This offset wasn't constant in the
four months that the observations lasted, but decreased progressively
from 33 to 26 million kilometres. In other words, the Earth-Jupiter
distance computed with the circular orbits introduced a discrepancy, in
the increasing distance of the two planets, of about 7 million km,
corresponding to 23 seconds at the speed of light, compared to the `real
word' situation given by VSOP87 ephemeris, in the time span of the
observational period of four months.

\thispagestyle{fancy}
To compute the average time between two emersions, which we call Io's
average synodic period, we started from the sidereal periods of
Jupiter's orbit around the Sun and of Io's around Jupiter. Dividing
these two periods we obtain the number \emph{N} of Io's orbits in a
Jovian year (\emph{N}=2449.3778). To `catch up' with the Jupiter shadow
Io loses one revolution around Jupiter in one Jovian year, and the
synodic period of Io is therefore obtained from its sidereal period
multiplied by the ratio between the number \emph{N} of its orbits in a
Jovian year divided by \emph{N-1} (see Table 4). The synodic period of
Io is analogous to the synodic period of our Moon, i.e., the lunar
month, which is the time between successive full (or new) moons. Every
time Io is full as seen from Jupiter there is an eclipse, while for our
moon eclipses are the exception due to the high inclination of the plane
of the lunar orbit with respect to the ecliptic (about 5°) and the fact
that Earth's shadow is much smaller than Jupiter's. Curiously, our Moon
and Io have about the same dimensions in terms of their physical sizes
and orbits around their respective planets.

\par\medskip
\noindent\begin{minipage}{\columnwidth}
\begingroup
\fontsize{6.65}{8.05}\selectfont
\setlength{\tabcolsep}{2.2pt}
{\sffamily\fontsize{7.2}{8.7}\selectfont\color{JournalGray}\RaggedRight
Table 4. Io's orbital periods used in this work.\par}
\nopagebreak[4]\vspace{2.5pt}
\begin{tabularx}{\columnwidth}{@{}
  >{\RaggedRight\arraybackslash}X
  >{\centering\arraybackslash}p{15.5mm}
  >{\centering\arraybackslash}p{17mm}@{}}
\toprule
\multicolumn{3}{c}{Io's Period} \\
Type & (days) & (seconds) \\
\midrule
Sidereal & 1.769137786 & 152853.505 \\
Synodic (average in one Jovian year) & 1.769860313 & 152915.931 \\
Synodic (correction for Jovian elliptic orbit) & 1.769921875 &
152921.247 \\
Synodic (computed from ephemeris in observations' timespan) &
1.769876150 & 152917.094 \\
\bottomrule
\end{tabularx}
\endgroup
\end{minipage}
\par\medskip

The difference between the sidereal and synodic periods of Io are easily
understood by looking at panels b and c of Figure 1. The different
inclinations of Jupiter's shadow are due to the movement of the planet
in its orbit around the Sun during the time Io orbited 70 times around
Jupiter. In exactly one sidereal period (and, consequently, also in 70
orbital periods), Io arrives in the same orbital position relative to
the stars (e.g., at exactly the 12 o'clock position in the black
circle), but not relative to the shadow. In fact, in this time the
shadow moved and it is slightly ahead of the satellite.

Let's start counting the time of all the observations from the instant
of the first emersion. Then, for each successive time of emersion,
calculate the time interval since this zero point. Dividing this
interval by the synodic period of Io, a number close to an integer will
result, corresponding to the number of Io's orbits around Juiter.

The decimal part of this number represents the delay introduced by the
finite speed of light. It indicates the time---expressed as a fraction
of the synodic period---required for light to reach Earth as it moves
away from Jupiter (assuming observations, as ours, are made after
Jupiter\textquotesingle s opposition; the reverse case, an advance,
would occur if observations were made before opposition, when Earth is
approaching Jupiter). Multiplying this fraction by the synodic period we
obtain the time delay of the light, Δ \emph{t}. Knowing the increased
Earth-Jupiter distance Δ \emph{x} from the first observation, we can
calculate the speed of light.

We put in a graph the distances Δ \emph{x} in function of the time delay
Δ \emph{t} since the first observed emersion (Figure 2). For the three
series of data, with linear regressions we found the slopes of the lines
that give the speed of light. Results are summarized in Table 5. It is
evident observing the standard errors of the slopes and
R\textsuperscript{2} values that the series fit very well with straight
lines. The found velocities are also very close each other, even if
there is not a range in which all of the confidence intervals overlap.

\subsection{Estimation of the observing errors}

We found that the standard deviation of a single measurement of time of
emersion is, using eye observations, 12-13 s. These values were obtained
from the regression lines (see eq. 8.15 in Taylor~\journalcite{27}), and are compatible to
what is estimated by the observers. The error in time detection is about
three times lower using the digital data, as we can expect. Very similar
results are obtained for the standard deviations of the differences
between observed times and the timing obtained with Stellarium (standard
deviations of 14 and 15 seconds for eye observations and 3 seconds for
digital camera data; see last line of Table 2).

In order to find out the main sources of errors in eye observations, we
used the longer series of FF that includes 8 successful observations. Of
these, 4 were obtained in good to optimal conditions of transparency and
seeing\footnote{\emph{Seeing} is the term used by astronomers to refer to the
  atmospheric turbulence that smears and blurs images of celestial
  objects.}. The other four were affected by atmospheric hazes and
veils (three observations) or by very poor seeing and pointing issue
(one observation).In fact, by manually adjusting the
telescope\textquotesingle s aim to recentre Jupiter just before Io's
emersion, on this occasion, once the vibrations had subsided a few
seconds later, the satellite was already visible. This effectively
delayed the detection of its emergence by a few seconds. Splitting the
single time series of FF into two groups (optimal conditions and not)
gives differences with Stellarium of -5 seconds (± 5 s) and +17 seconds
(± 9 s), respectively. This shows that the atmospheric conditions are
the main contributor to the total error in eye timings. In fact, in
optimal conditions the standard error is reduced substantially.
Moreover, the non-optimal observations are delayed by about 20 seconds,
compared to the optimal four. We also checked the naked eye limiting
magnitude and the sky brightness with a Sky Quality Meter (SQM), and we
found that on the three veiled night the sky was three to six times
brighter than during the other observations due to the diffusion by the
veils and haze of light pollution and moonlight, when present. This may
have contributed to the above-mentioned delay in detection.

The errors due to uncertainties in the time apps and radio-controlled
watches are well below 0.1 s and the uncertainty in transferring
manually the time to a stopwatch is below 0.5 s (otherwise the
synchronization was repeated).

\end{multicols}\vfill\thispagestyle{fancy}\begin{widetable}
\noindent\begin{minipage}{\textwidth}
\begin{journalcaption}

Table 5. Results of speed of light calculated with a uniform circular
model

\end{journalcaption}

\vspace{1.5pt}
\begin{tabularx}{\textwidth}{@{}
  >{\RaggedRight\arraybackslash}p{0.18\textwidth}
  >{\RaggedRight\arraybackslash}p{0.18\textwidth}
  >{\RaggedRight\arraybackslash}p{0.23\textwidth}
  >{\centering\arraybackslash}p{0.08\textwidth}
  >{\RaggedRight\arraybackslash}X@{}}
\toprule
Observer & Speed of light

(km s\textsuperscript{-1}) & Speed standard deviation

(km s\textsuperscript{-1}) & R\textsuperscript{2} & Standard deviation
in the timing of a single observation~(s) \\
\midrule
Fabio Falchi (FF) & 258800 & 4400 & 0.9982 & 13 \\
Paolo Gattillo (PG) & 261900 & 8200 & 0.9981 & 12 \\
Riccardo Furgoni (RF) & 267670 & 890 & 0.99998 & 4 \\
Weighted average & 267270 & 870 & & \\
\bottomrule
\end{tabularx}
\end{minipage}

\end{widetable}\clearpage\begin{multicols}{2}

\end{multicols}\begin{widefigure}

\includegraphics[width=17cm,height=16.928cm]{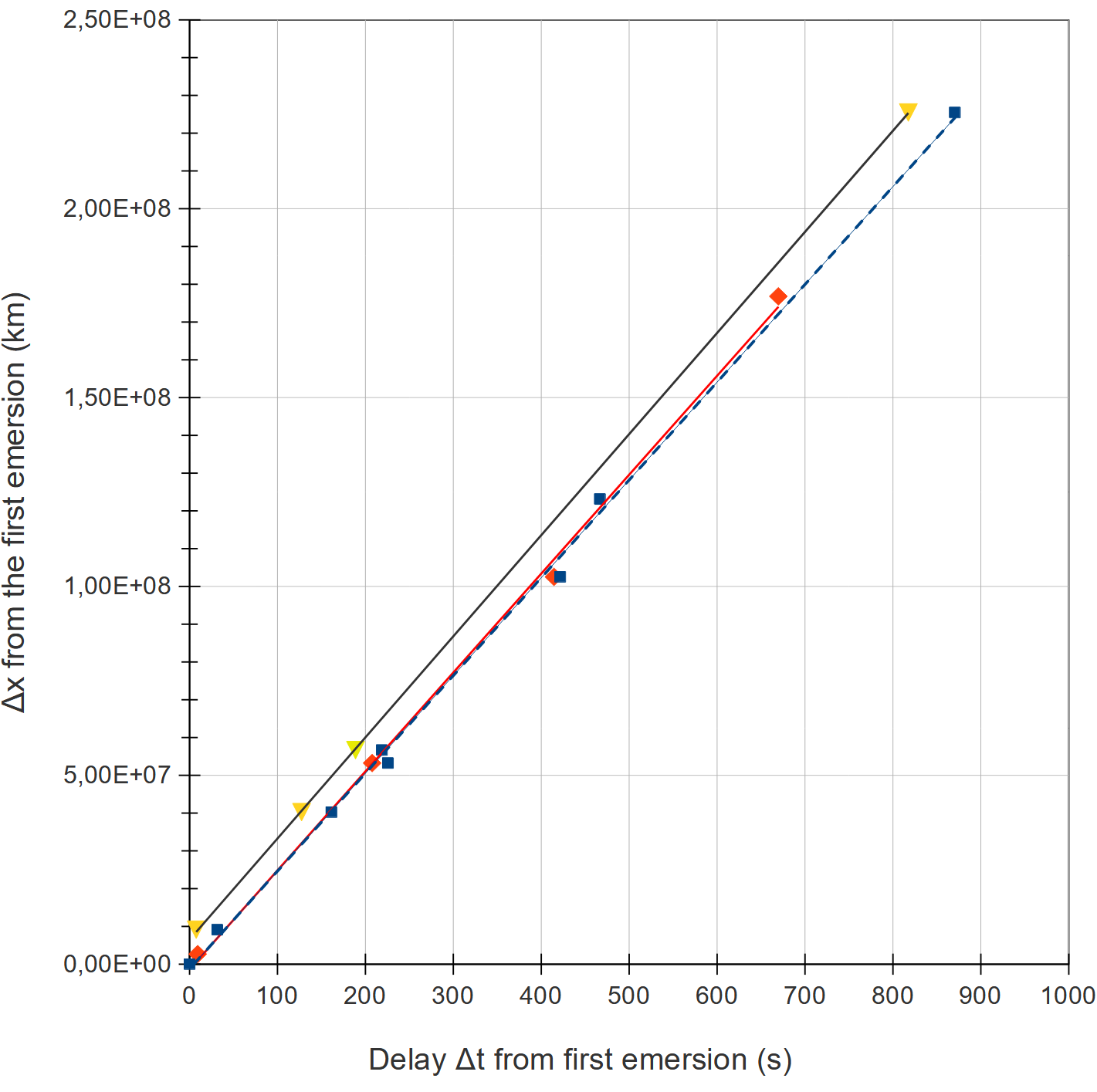}

\begin{journalcaption}

Figure 2. The three series of observations where the symbols represent
the increase in Jupiter-Earth distance from the first emersion in
function of time delay in emersions. Blue squares are FF's observations
(regression line blue dashed), red diamonds are PG's (regression line
solid red) and yellow triangles RF's (regression line solid black). The
slopes of the lines of regressions, almost identical for the three
series, indicate the speed of light.

\end{journalcaption}\end{widefigure}\begin{multicols}{2}

As the observations are judged to be very good, we need to explain the
difference of 11-14\% between the correct value of \emph{c} (speed of
light=299792 km s\textsuperscript{-1}) and our measured values.
Adjusting the Earth-Jupiter distances to the values given by modern
ephemeris with Stellarium, instead of our uniform circular model,
improves our \emph{c} value by 3-4\%. Most of the difference remains to
be explained.

\subsection{Elliptical orbits model}

The first thing we tried was to introduce a correction to take into
account the variable speed of Jupiter in its orbit around the Sun. In
fact, in the period of our observations Jupiter was still very close to
its perihelion, so both its linear speed and angular speed were higher
than those we adopted in the simple uniform circular motions we used in
our first analysis. Near perihelion the speed of the Jupiter system is
higher, while the sidereal period of Io remains unchanged (except for
reciprocal interactions with the other Galilean satellites, here
disregarded). This means that the orientation of Jupiter's shadow change
faster than on average and the synodic period of Io increases, as the
satellite has to travel a longer track than usual to `run after' the
shadow.

To calculate the synodic period of Io with this correction we operated
in this way:

- we found the heliographic inertial longitude of Jupiter\footnote{Taken from: https://omniweb.gsfc.nasa.gov/coho/helios/heli.html} at the times of first and last observed emersion;

- the difference between these longitudes gives the angle
\emph{a}\textsubscript{\emph{elliptical}} Jupiter travelled from the
first and last eclipse;

- we calculate the angle \emph{a}\textsubscript{\emph{uniform}} using
the constant angular speed of Jupiter from the uniform circular model;

- D\emph{a =} \emph{a}\textsubscript{\emph{elliptical}}\emph{
-a}\textsubscript{\emph{uniform}} gives the additional angle that Io
should cover to `run after' Jupiter's shadow;

- D\emph{a }divided by the sidereal angular velocity of Io,
\emph{w}\textsubscript{\emph{I}\emph{o}} (=2.355x10\textsuperscript{-3}
°/s), gives the additional time accumulated in 70 orbits,
D\emph{t}\textsubscript{\emph{70}}=372.1 s, needed to run after the
shadow, which is equivalent to D\emph{t}=5.32 s of additional time per
orbit;

- summing this D\emph{t} to the average synodic period gives the
corrected synodic period \emph{T}\textsubscript{synodic corr} =
152921.25 s.

Calculating the D\emph{t} using the orbital velocity of Jupiter and its
distance from the Sun given by Stellarium in the middle of the range of
observation times, i.e., on January 9, 2024, gives a similar result of
D\emph{t}=5.48 s.

Unfortunately, this over-corrects too much, giving \emph{c} values,
depending on the three observers, from 441000 km s\textsuperscript{-1}
to 460000 km s\textsuperscript{-1}, corresponding 1 A.U. light-travel
times of about 5.5 minutes. It is evident that other factors are at
work. These additional factors have already been analysed well by James
Appleton~\journalcite{28}. He found that the main correction
factor is indeed the ellipticity of Jupiter's orbit. Not taking into
account this correction may give calculated values of \emph{c} very
different from the real value, from 150000 km s\textsuperscript{-1
}(i.e. 1/c= 16 minutes to travel 1 A.U.) to, in very rare cases (about
10 times in 600 years, as seen in Fig 9 in Appleton), when the eclipses
anticipates when instead they should be retarded by Earth receding from
Jupiter (note: 1/\emph{c} in this case would be negative; 1/\emph{c}=0
means an infinite speed of light). Other factors influencing the
variability of Io's orbital motion are the resonance between Io and its
closest large satellites, Europa and Ganymede and, to a lower degree,
the inclination of Io's orbit.

The correction for the ellipticity of Jupiter's orbit gives times to
cross 1 A.U., when the planet is near perihelion, from 4 to 5.6 minutes,
depending on the perihelion considered (about 5.3 minutes for the
perihelion of 2023), as shown in Figure 10 in Appleton. This is
compatible to our value of velocity --- calculated with ellipticity
corrections --- that correspond to 5.5 minutes to travel 1 A.U.

In our case, accounting for the eccentricity of
Jupiter\textquotesingle s orbit worsened the value obtained for the
speed of light, because our observations happened to coincide with a
period when the time required to traverse 1 AU calculated using the
uniform model, slightly exceeding 9 minutes, was already very close to
the actual 8.3 minutes required. Observations made in other periods
during the Jovian year may give much different results, and the
calculation of the speed of light made taking into account the
ellipticity of Jupiter's orbit may give a better estimation of \emph{c}
compared to the simple circular model.

\subsection{Variability of Io's synodic period}

We used the Stellarium predictions to calculate the synodic period of Io
during the 2006-2030 period, covering approximately two Jovian sidereal
years. The synodic period of Io was averaged for each corresponding 70
eclipses during each Jovian synodic year (i.e., the time from one
opposition to the next, when the Sun, Earth and Jupiter are
approximately on the same line). The results are shown in Figure 3.
Superimposed are the values of the average of these synodic periods
(solid line) and the simple synodic period obtained from uniform
circular orbits (dotted line). The approximately sinusoidal variability
of Io's synodic period during the Jupiter year due mainly to the
ellipticity of the orbit was already taken into account by Cassini since
1668~\journalcite{29}.

\end{multicols}\begin{widefigure}

\includegraphics[width=15cm,height=12.291cm]{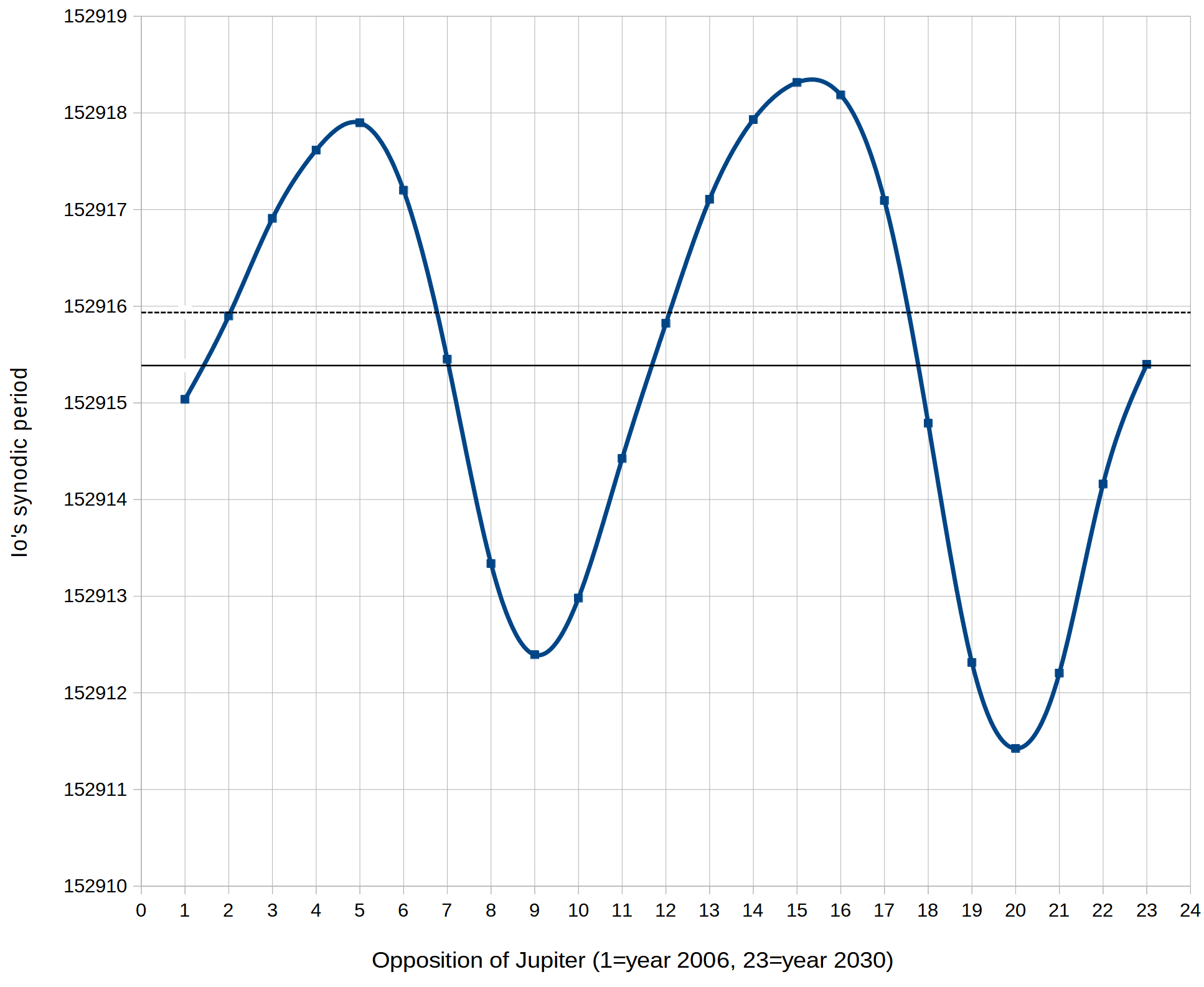}

\begin{journalcaption}

Figure 3. Synodic period of Io variability during 23 opposition of
Jupiter, from 2006 to 2030. The synodic period corresponding to our
observations is that of number 17 in the \emph{X} axis. Solid line
indicates the average of all these periods; dotted line indicates the
circular orbits' synodic period.

\end{journalcaption}\end{widefigure}\begin{multicols}{2}

\subsection{Calculation knowing the `exact' value of the synodic period}

To test how good the observations by eye and with a digital camera are,
we now presume to know the speed of light and the exact positions of
Jupiter (or Io) and Earth using the ephemeris given by Stellarium. In
this way, observing and timing the ends of eclipses, we can calculate
when Io emerged from Jupiter's shadow by subtracting the times needed
for light to travel from Io to Earth. These are the local (Jupiter)
times of emersions.

Using the first (November 7, 2023) and last (March 10, 2024) local times
of emersions, and dividing this time interval by the 70 orbits that Io
travelled in between, we obtained for the synodic period of Io a value
\emph{t}\textsubscript{\emph{ephemeris}}= 152917.1 s. We then used this
\emph{t}\textsubscript{\emph{ephemeris}} period to calculate the speed
of light with the three series of observations, by taking, as seen in
the uniform circular model analysis, the slopes of the regression lines.
The graphs are shown in Figure 4. The values of the speed of light are
summarized in Table 6.

\end{multicols}\clearpage\begin{widefigure}

\includegraphics[width=11.8cm,height=\textheight]{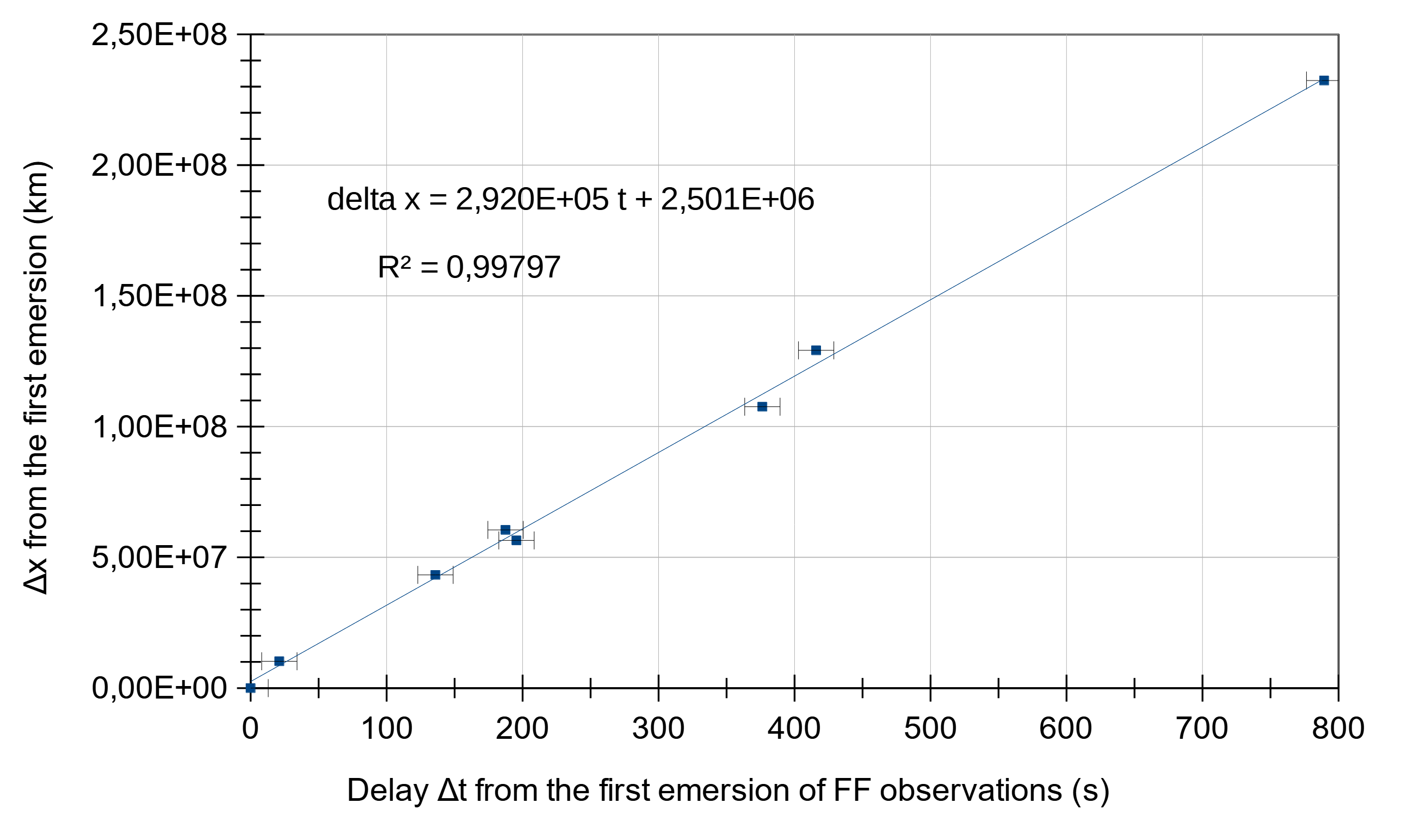}\\
\includegraphics[width=11.8cm,height=\textheight]{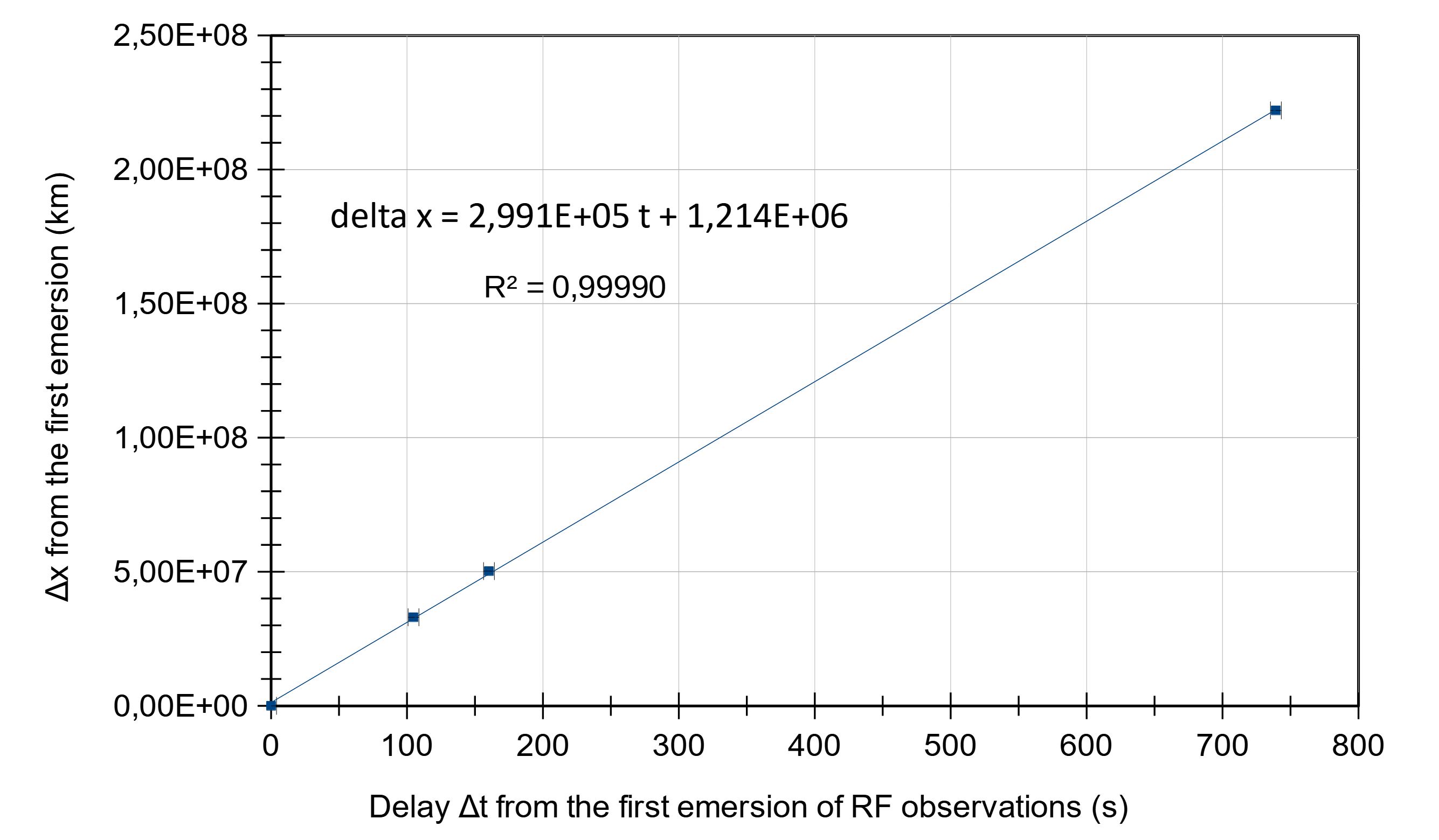}\\
\includegraphics[width=11.8cm,height=\textheight]{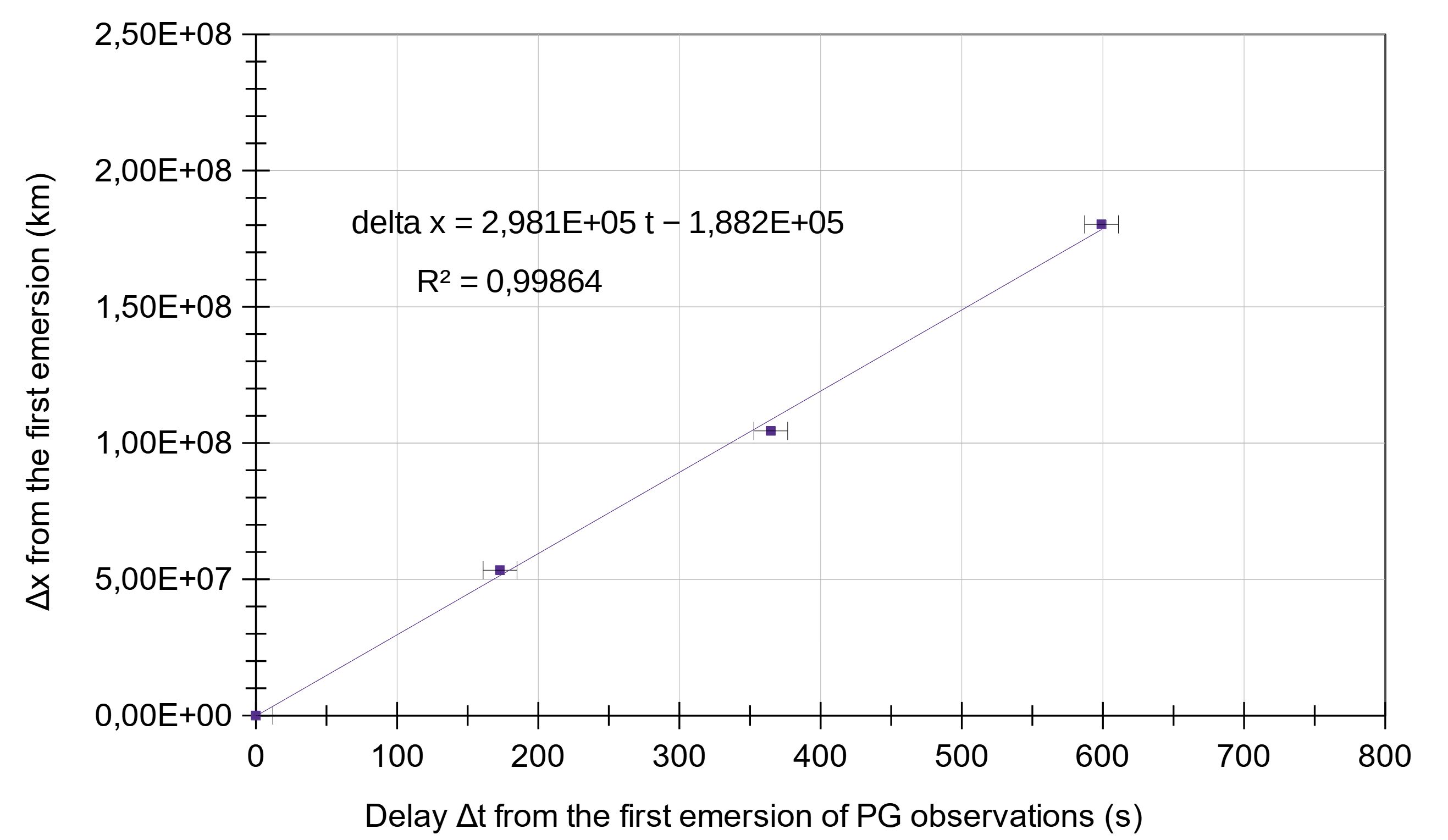}

\begin{journalcaption}

Figure 4. The three series of observations where the data-points
represent the increase in Jupiter-Earth distance from the first emersion
of each series in function of time delay D \emph{t }(note: in Figure 2,
differently from here, the zero of time was the first observation of
FF). From top to bottom: FF, RF, PG.

\end{journalcaption}\end{widefigure}\vfill\begin{widetable}

\begin{journalcaption}

Table 6. The speed of light as resulted from authors' time observations
of Io's end of eclipses applied to modern ephemerides.

\end{journalcaption}

\begin{longtable}[]{@{}
  >{\raggedright\arraybackslash}p{(\columnwidth - 6\tabcolsep) * \real{0.2500}}
  >{\raggedright\arraybackslash}p{(\columnwidth - 6\tabcolsep) * \real{0.2500}}
  >{\raggedright\arraybackslash}p{(\columnwidth - 6\tabcolsep) * \real{0.2500}}
  >{\raggedright\arraybackslash}p{(\columnwidth - 6\tabcolsep) * \real{0.2500}}@{}}
\toprule\noalign{}
Observer & Speed of light

(km s\textsuperscript{-1}) & Standard deviation

(km s\textsuperscript{-1}) & R\textsuperscript{2} \\
\midrule\noalign{}
\endhead
\bottomrule\noalign{}
\endlastfoot
Fabio Falchi (FF) & 292000 & 5400 & 0.99797 \\
Paolo Gattillo (PG) & 298100 & 7800 & 0.99864 \\
Riccardo Furgoni (RF) & 299100 & 2100 & 0.99990 \\
Weighted average & 298200 & 1900 & \\
\end{longtable}

\end{widetable}\clearpage\begin{multicols}{2}

The results of this analysis show that the observations by eye, when
combined, give values of speed of light differing about 2\% from the
correct value, while the observations with a digital camera differ by
only 0.2\%. The observations by eye are equivalent to those possible at
the time of Roemer, as we choose refracting telescopes with similar
diameter to those used by him and his colleagues. Against the precision
of Roemer's observations there was the accuracy of the clocks then
available, approximately 15 s/day, that could have been realistically
reduced to a few seconds if synchronized with the transits of stars in
the same nights of the observations of eclipses. As we found that the
uncertainty of a single observation by eye is about 15 seconds, the
uncertainty in Roemer's clocks should not have impacted heavily on his
observations.

\section{\texorpdfstring{IV DIDACTIC VALUE AND CONCLUSIONS
}{IV DIDACTIC VALUE AND CONCLUSIONS }}

We found that Roemer's method to determine the speed of light can be
repeated easily using very small telescopes (80-120 mm diameters) and
visual observations at the eyepiece. The experimental errors due to the
uncertainty in the determination of the exact timing of the ends of
eclipses are small and influence negligibly the calculation of the speed
of light. Even using a very simple uniform circular-motion model can
give good estimations of the speed of light, circa 260000 km
s\textsuperscript{-1} in our case, but this may change substantially
with the year of observation. We also found that using a more accurate
elliptical orbit model does not necessarily bring \emph{c} values closer
to the real one, in our case giving in fact about 450000 km
s\textsuperscript{-1}. This is due to the other factors influencing the
revolution period of Io around Jupiter. Using, at last, the synodic
period calculated with modern ephemerides allowed us to reach a \emph{c}
value of 298200±1900 km s\textsuperscript{-1}, which is very close to
the modern value.

Our findings demonstrate that Roemer's method, beyond being
scientifically sound, was also operatively and practically viable to
prove that light propagates at finite speed. To obtain more accurate
values compared to those calculated by Huygens using Roemer's data, it
would have been necessary to have a deeper knowledge of Io's ephemerides
not possible in Roemer's time. In fact, the values of \emph{c} obtained
in different years using uniform circular motion or a correction for the
ellipticity of Jupiter's orbit and its variable speed along it may
differ substantially. Probably for this reason, Roemer was reluctant ---
as far as we know --- in giving a value for the speed of light and
limited himself in declaring only its finite speed.

All this can be of great didactic value, particularly when integrated
into undergraduate physics and history of science curricula, helping
students to formulate hypotheses on the causes of the retardment or
anticipation of the periods of revolution of Io during each period of
visibility of the Jupiter system and testing them inside different
theoretical frames of increasing complexity.

In the context of physics laboratory, the project provides a rigorous
framework for students to engage with real-world data acquisition and
error analysis. Beyond the simple calculation of a physical constant,
students are required to confront human versus instrumental
observations. By comparing manual timings with CMOS or CCD sensor
recordings and observations made in different atmospheric conditions,
they gain a practical understanding of systematic and random errors.
Furthermore, the transition from a basic circular orbit model to more
advanced mathematical descriptions encourages students to evaluate the
trade-offs between model complexity and predictive accuracy, a
fundamental skill in scientific research.

For history of science courses, this experiment serves as a powerful
"hands-on" pedagogical tool. Of great didactic importance is the
possibility to read what the scientists at the forefront of knowledge in
17th century wrote in their communications to academies of science and
in their correspondence with each other. Moreover, by using telescopes
with apertures and resolutions comparable to those available in the late
17th century, students can directly experience some of the challenges
faced by Roemer and his contemporaries. This immersion facilitates a
deeper epistemological discussion on how scientific paradigms shift;
students witness first-hand how anomalies in the observed orbital period
of Io led to the groundbreaking realization that light propagates at a
finite speed. This approach transforms the history of physics from a
series of dates and names into a living process of discovery, allowing
students to replicate a milestone of human thought using both
period-appropriate logic and modern verification tools like Stellarium
and digital cameras.

To ensure these scenarios are easily replicable in a university setting,
we have provided a detailed description of the experimental setup used
in our campaign. Successful implementation requires relatively
accessible equipment, such as a standard 8-inch Schmidt-Cassegrain
telescope or a small refractor, paired with a digital camera or, simply,
the eye at the eyepiece. We emphasize the importance of time
synchronization---e.g. using Network Time Protocol (NTP) servers---to
ensure that student observations can be meaningfully compared with
modern ephemerides. By following this structured approach, academic
institutions can offer a complete educational experience that
encompasses observational technique, historical analysis, and
contemporary computational verification.

\section{ACKNOWLEDGEMENTS}

This work is dedicated to our co-author Maurizio, who passed away before
the manuscript was completed.\\
We would like to thank the three anonymous reviewers, whose comments,
critics and suggestions helped improving this paper.

Special thanks to Dr. John Barentine, who kindly helped to improve some
passages of the manuscript.

\section{REFERENCES}

\begin{journalreferences}
\item[\hypertarget{ref-1}{\hyperlink{cite-1}{\textcolor{JournalText}{\textbf{[1]}}}}] M. Francesio, A. Lai, and G. Zanini, ``De mora luminis, sive
  diffusione luminis successiva -- Ricerca sulle prime misure della
  velocità della luce,'' La Fisica nella Scuola \textbf{XII}(3), 68--82
  (1979).
\item[\hypertarget{ref-2}{\hyperlink{cite-2}{\textcolor{JournalText}{\textbf{[2]}}}}] Galilei, Galileo, \emph{Discorsi e Dimostrazioni
  Matematiche, Intorno a Due Nuove Scienze} (appresso gli Elsevirii,
  Leida, 1638).
\item[\hypertarget{ref-3}{\hyperlink{cite-3}{\textcolor{JournalText}{\textbf{[3]}}}}] R. Foschi, and M.
  Leone, ``Galileo, Measurement of the Velocity of Light, and the
  Reaction Times,'' Perception \textbf{38}(8), 1251--1259 (2009).
\item[\hypertarget{ref-4}{\hyperlink{cite-4}{\textcolor{JournalText}{\textbf{[4]}}}}] C. Huygens,
  \emph{Oeuvres Complètes. Tome VIII. Correspondance 1676-1684},
  Johannes Bosscha jr. ed. (Martinus Nijhoff, Den Haag, 1899), pp.30-31,
  letter n.2103, dated September 16, 1677.
\item[\hypertarget{ref-5}{\hyperlink{cite-5}{\textcolor{JournalText}{\textbf{[5]}}}}] ``A
  demonstration concerning the motion of light, communicated from Paris,
  in the journal des scavans, and here made English,'' Phil. Trans. R.
  Soc. \textbf{12}(136), 893--894 (1677).
\item[\hypertarget{ref-6}{\hyperlink{cite-6}{\textcolor{JournalText}{\textbf{[6]}}}}] S.A. Bedini, ``Lens Making for Scientific
  Instrumentation in the Seventeenth Century,'' Appl. Opt.
  \textbf{5}(5), 687 (1966).
\item[\hypertarget{ref-7}{\hyperlink{cite-7}{\textcolor{JournalText}{\textbf{[7]}}}}] M. Miniati,
  A. Van Helden, V. Greco, and G. Molesini, ``Seventeenth-century
  telescope optics of Torricelli, Divini, and Campani,'' Appl. Opt.
  \textbf{41}(4), 644 (2002).
\item[\hypertarget{ref-8}{\hyperlink{cite-8}{\textcolor{JournalText}{\textbf{[8]}}}}] J.D.
  Cassini, ``An extract of Signor Cassini's letter concerning a spot
  lately seen in the sun; together with a remarkable observation of
  Saturn, made by the same,'' Phil. Trans. R. Soc. \textbf{11}(128),
  689--690 (1676).
\item[\hypertarget{ref-9}{\hyperlink{cite-9}{\textcolor{JournalText}{\textbf{[9]}}}}] J. Lozi,
  J.-M. Reess, A. Semery, E. Lhomé, S. Jacquinod, M. Combes, P.
  Bernardi, R. Andretta, M. Motisi, L. Bobis, and E. Kaftan, ``Could
  Jean-Dominique Cassini see the famous division in Saturn's rings?,''
  in \emph{SPIE Proceedings}, edited by S. Shaklan, (SPIE, San Diego,
  California, United States, 2013), p. 88641M.
\item[\hypertarget{ref-10}{\hyperlink{cite-10}{\textcolor{JournalText}{\textbf{[10]}}}}] A.R. Willms,
  P.M. Kitanov, and W.F. Langford, ``Huygens' clocks revisited,'' R.
  Soc. Open Sci. \textbf{4}(9), 170777 (2017).
\item[\hypertarget{ref-11}{\hyperlink{cite-11}{\textcolor{JournalText}{\textbf{[11]}}}}] J. Picard, \emph{Mesure de La Terre} (Imprimerie
  royale, Paris, 1671).
\item[\hypertarget{ref-12}{\hyperlink{cite-12}{\textcolor{JournalText}{\textbf{[12]}}}}] F. Hultsch, \emph{Griechische und römische metrologie}
  (Weidmann, Berlin, 1882), p.61.
\item[\hypertarget{ref-13}{\hyperlink{cite-13}{\textcolor{JournalText}{\textbf{[13]}}}}] L.
  Russo, ``Ptolemy's longitudes and Eratosthenes' measurement of the
  earth's circumference,'' Math. Mech. Compl. Sys. \textbf{1}(1), 67--79
  (2013).
\item[\hypertarget{ref-14}{\hyperlink{cite-14}{\textcolor{JournalText}{\textbf{[14]}}}}] L. Russo, ``Far-reaching
  Hellenistic geographical knowledge hidden in Ptolemy's data,'' Math.
  Mech. Compl. Sys. \textbf{6}(3), 181--200 (2018).
\item[\hypertarget{ref-15}{\hyperlink{cite-15}{\textcolor{JournalText}{\textbf{[15]}}}}] Favaro, Antonio, \emph{Le Opere Di Galileo Galilei} (Barbera, Firenze,
  1932), Vol. V, pp. 423-425.
\item[\hypertarget{ref-16}{\hyperlink{cite-16}{\textcolor{JournalText}{\textbf{[16]}}}}] Favaro, Antonio, \emph{Le Opere Di Galileo Galilei} (Barbera, Firenze,
  1934), Vol. XII, p. 291-295.
\item[\hypertarget{ref-17}{\hyperlink{cite-17}{\textcolor{JournalText}{\textbf{[17]}}}}] Favaro, Antonio, \emph{Le Opere Di Galileo Galilei} (Barbera, Firenze,
  1932), Vol. V, p. 415-418.
\item[\hypertarget{ref-18}{\hyperlink{cite-18}{\textcolor{JournalText}{\textbf{[18]}}}}] O.
  Roemer, ``Démonstration touchant le mouvement de la lumière trouvé par
  M. Roemer de l'Académie des sciences,'' Journal Des Sçavans, 233--236
  (1676).
\item[\hypertarget{ref-19}{\hyperlink{cite-19}{\textcolor{JournalText}{\textbf{[19]}}}}] L. Bobis, and J.
  Lequeux, ``Cassini, Rømer, and the velocity of light,'' Journal of
  Astronomical History and Heritage \textbf{11}, 97--105 (2008).
\item[\hypertarget{ref-20}{\hyperlink{cite-20}{\textcolor{JournalText}{\textbf{[20]}}}}] J.-D.
  Cassini, \emph{Les Hypothèses et Les Tables Des Satellites de Jupiter,
  Réformées Sur de Nouvelles Observations} (Imprimerie royale, Paris,
  1693) p.52.
\item[\hypertarget{ref-21}{\hyperlink{cite-21}{\textcolor{JournalText}{\textbf{[21]}}}}] C. Huygens,
  \emph{Oeuvres Complètes. Tome VIII. Correspondance 1676-1684},
  Johannes Bosscha jr. (Martinus Nijhoff, Den Haag, 1899) pp.33-35,
  Letter 2104, dated September 30, 1677.
\item[\hypertarget{ref-22}{\hyperlink{cite-22}{\textcolor{JournalText}{\textbf{[22]}}}}] J.H. Shea, ``Ole Rømer, the speed of light, the
  apparent period of Io, the Doppler effect, and the dynamics of Earth
  and Jupiter,'' American Journal of Physics \textbf{66}(7), 561--569
  (1998).
\item[\hypertarget{ref-23}{\hyperlink{cite-23}{\textcolor{JournalText}{\textbf{[23]}}}}] J. Mira-Pérez, and S.X. Bará,
  ``Determining Longitude: A Brief History,'' Physics Today
  \textbf{58}(10), 15--16 (2005).
\item[\hypertarget{ref-24}{\hyperlink{cite-24}{\textcolor{JournalText}{\textbf{[24]}}}}] O. Römer, ``Confirmatio doctrinae de Mora Luminis Ex
  novis Observationibus anni 1677,'' C. Huygens, Oeuvres Complètes. Tome
  VIII. Correspondance 1676-1684, Johannes Bosscha Jr. Ed. (Martinus
  Nijhoff, Den Haag, 1899), Pp.56-58 (No 2116.), (n.d.).
\item[\hypertarget{ref-25}{\hyperlink{cite-25}{\textcolor{JournalText}{\textbf{[25]}}}}] A.
  Mallama, C. Stockdale, B.A. Krobusek, and P. Nelson, ``Assessment of
  the resonant perturbation errors in Galilean satellite ephemerides
  using precisely measured eclipse times,'' Icarus \textbf{210}(1),
  346--357 (2010).
\item[\hypertarget{ref-26}{\hyperlink{cite-26}{\textcolor{JournalText}{\textbf{[26]}}}}] P. Bretagnon, and G. Francou, ``Planetary Theories
  in rectangular and spherical variables: VSOP87 solution.,'' Astronomy
  and Astrophysics \textbf{202}, 309 (1988).
\item[\hypertarget{ref-27}{\hyperlink{cite-27}{\textcolor{JournalText}{\textbf{[27]}}}}] J. Taylor,
  \emph{Introduction to Error Analysis, the Study of Uncertainties in
  Physical Measurements, 2nd Edition} (1997).
\item[\hypertarget{ref-28}{\hyperlink{cite-28}{\textcolor{JournalText}{\textbf{[28]}}}}] J. Appleton, ``Rømer revisited: A modern estimation
  of the speed of light from observations of Jupiter's Galilean
  satellites,'' Journal of the British Astronomical Association
  \textbf{126}, 139--148 (2016).
\item[\hypertarget{ref-29}{\hyperlink{cite-29}{\textcolor{JournalText}{\textbf{[29]}}}}] Cassini, Giovanni Domenico, \emph{Ephemerides Bononienses
  Mediceorum syderum} (Typis Emilij Mariae \& fratrum de Manolessijs,
  Bologna, 1668).
\end{journalreferences}
\end{multicols}
\end{document}